\documentstyle[12pt]{article}

\def\beq{\begin{equation}}
\def\eeq{\end{equation}}
\def\bce{\begin{center}}
\def\ece{\end{center}}
\def\bea{\begin{eqnarray}}
\def\eea{\end{eqnarray}}
\def\ben{\begin{enumerate}}
\def\een{\end{enumerate}}

\def\ni{\noindent}
\def\nn{\nonumber}
\def\bs{\bigskip}
\def\ms{\medskip}
\def\wt{\widetilde}

\def\tr{\mbox{Tr}\, }
\def\brr{\begin{array}}
\def\err{\end{array}}

\begin{document}

\hfill CEAB 95/1207

\hfill December 1995

\hfill hep-th/9601101


\begin{center}

{\LARGE \bf
The renormalization group and spontaneous compactification of a
higher-dimensional scalar field theory in curved spacetime}

\vspace{5mm}

{\sc E. Elizalde}\footnote{E-mail: eli@ecm.ub.es} \\
Center for Advanced Studies CEAB, CSIC, Cam\'{\i} de Santa
B\`arbara,
17300 Blanes,
\\ and Department ECM and IFAE, Faculty of Physics,
University of Barcelona, \\ Diagonal 647, 08028 Barcelona,
 Spain \\
{\sc R. Kantowski} \footnote{E-mail: kantowski@phyast.nhn.uoknor.edu} \\
Department of Physics and Astronomy,
University of Oklahoma, \\
Norman, OK 73019, USA
\\
 and \\
{\sc S.D. Odintsov}\footnote{E-mail: sergei@ecm.ub.es.}
\\
 Tomsk Pedagogical University, 634041 Tomsk, Russia, \\
and Department ECM, Faculty of Physics,
University of  Barcelona, \\  Diagonal 647, 08028 Barcelona,
Spain \\

\vspace{12mm}

{\bf Abstract}
\end{center}

The renormalization group (RG) is used to study the
asymptotically free $\phi_6^3$-theory in curved
spacetime. Several forms of the
RG equations for the effective potential are
formulated. By solving these equations we obtain the one-loop
effective potential as well as its explicit forms in  the case
of strong gravitational fields and strong scalar fields. Using
zeta function techniques, the one-loop and
corresponding RG
improved vacuum energies are found for the Kaluza-Klein backgrounds
$R^4\times S^1\times S^1$ and $R^4\times S^2$.
They are given in terms of exponentially convergent series,
appropriate for numerical calculations. A study of these vacuum energies
as a function of compactification lengths and other couplings
shows that spontaneous compactification
can be qualitatively different when the RG improved energy is used.

\vfill

\newpage

\bce
{\large \bf I. \ Introduction.}
\ece

The renormalization group (RG) has long been used to ``improve" loop
corrections in perturbative quantum field theory.
Gell-Mann and Low \cite{GML}
first used it to study the asymptotic behavior of Green's functions and
in the classic work of Coleman and Weinberg \cite{22} the RG equation was
used to improve the effective potential and study spontaneous symmetry breaking.
This is just one of the many different applications that the RG has had
in quantum filed theory.
Recently it has been employed to put lower limits on the Higg's mass of
the standard model \cite{23}. In this paper we put it to use in a
renormalizable Kaluza-Klein
model, arguing that RG improvments are necessary if stability of the
internal dimensions
are to be correctly predicted. We will also develop the RG technique in
order to study the scalar effective potential in this model.

We start in Sec. II using the RG equations to arrive at the 1-loop
effective potential starting from a classical $\phi^3$ scalar
field theory defined on a 6-dimensional curved space (denoted by $\phi^3_6$).
This is a
renormalizable theory which is coupled to the curvature tensor and its
square. We additionally use the RG equations to find the asymptotic
behavior of the effective potential when either the gravitational field
is strong or when the scalar field is intense. In Sec. III we give the
1-loop vacuum energies for this scalar field on backgrounds
$R^4\times S^1\times S^1$ and  $R^4\times S^2$. We then compute
RG improvements to these energies. We conclude that qualitative changes
have occurred, i.e., minima have disappeared from the vacuum energy and
that Kaluza-Klein stability will be correspondingly affected. In the
conclusions we also mention other possible applications of the RG
techniques in the context of Kaluza-Klein theories, as
higher-derivative theories and renormalizable theories in the ``modern
sense.''
 \bs

\bce
{\large \bf II. \ A renormalizable self-interacting scalar theory
 in D=6 curved spacetime.}
\ece

As an example of a renormalizable theory in higher-dimensional curved
spacetime, we consider the following action in $D=6$:
\bea
L &=& L_m + L_{ext}, \nn \\
L_m &=& -\frac{1}{2} \phi \Box \phi + \frac{1}{2} M^2 \phi^2 +
\frac{1}{3!} g \phi^3 + h\phi +  \frac{1}{2} \xi R \phi^2 +
\eta_1 R \phi \nn \\ && + \eta_3 R^2 \phi +  \eta_4 R_{\mu\nu}
R^{\mu\nu} \phi +  \eta_5 R_{\mu\nu\alpha\beta}
R^{\mu\nu\alpha\beta} \phi, \label{2.1} \\
L_{ext} &=& - \left( \Lambda + kR + \alpha_1
R_{\mu\nu\alpha\beta}^2  + \alpha_2 R_{\mu\nu}^2 + \alpha_3 R^2 +
\alpha_4 R^3 + \alpha_5 R R_{\mu\nu}^2 + \alpha_6 R
R_{\mu\nu\alpha\beta}^2 \right. \nn \\ && + \left. \alpha_7
R_{\mu\nu} R^\mu_{\ \sigma} R^{\nu\sigma} +  \alpha_8 R_{\mu\nu}
R_{\rho\sigma} R^{\mu\rho\nu\sigma} +\alpha_9 R_{\mu\nu}
R^{\mu\lambda\rho\sigma} R^\nu_{\ \lambda\rho\sigma} +
\alpha_{10} R_{\mu\nu\rho\sigma} R^{\mu\nu}_{\ \ \lambda\tau}
R^{\rho\sigma\lambda\tau} \right). \nn
\eea
Here $L_m$ and $L_{ext}$ are the Lagrangians of matter and
external fields, respectively, and $\phi$ is a scalar. The
Lagrangian (\ref{2.1}) represents the generalization to curved
space of a  renormalizable $\phi_6^3$ theory \cite{1}. Such a
theory in curved spacetime was considered a few years ago in Refs.~
\cite{2}-\cite{4}. Here the notation of Ref.~\cite{3} will be
adopted. In that reference, a one-loop analysis was carried out. The
form of $L_{ext}$ in
(\ref{2.1}), as well as of the non-minimal gravitational terms in
$L_m$, are such as to make the theory multiplicatively
renormalizable in curved spacetime. We will consider only spacetimes
of constant curvature, excluding  terms of the form
$\phi \Box R$, etc., from the Lagrangian (\ref{2.1}).
Finally, $\lambda_i =\{ M^2, g, h, \xi, \ldots, \alpha_{10} \}$
are all coupling constants whose dimensionality is clear from the
form of the Lagrangian (\ref{2.1}).

One-loop divergences of the model
(\ref{2.1}) are found in Ref.~\cite{3}. They yield the
following running coupling constants (we give here their explicit
expressions in the massless sector only):
\bea
g^2(t) &=& g^2B^{-1}(t), \qquad B(t)=1+ \frac{3g^2t}{2(4\pi)^3},
\nn \\
\xi (t)&=& \frac{1}{5} + \left( \xi - \frac{1}{5} \right) B^{-
5/9} (t), \nn \\
\eta_1(t) &=& \eta_1 B^{1/18}(t),
\label{2.2} \\
\eta_3(t) &=& B^{1/18}(t) \left[ \eta_3 - \frac{1}{1200g}  \left(
B^{4/9} (t) -1 \right) + \frac{1}{5g}  \left( \xi - \frac{1}{5}
\right) \left( B^{-1/9} (t) -1 \right) \right. \nn \\
&& \left. + \frac{1}{2g}  \left( \xi - \frac{1}{5} \right)^2
\left( B^{-2/3} (t) -1 \right) \right], \nn \\
\eta_{4,5}(t) &=& B^{1/18}(t) \left[ \eta_{4,5} \pm
\frac{1}{120g}  \left( B^{4/9} (t) -1 \right) \right], \nn\\
h(t)&=& h B^{1/18}(t).\nn
\eea
It is clear from expression (\ref{2.2}) that the theory is
asymptotically free at high energies ($g^2(t) \rightarrow 0$),
and that it is asymptotically conformal invariant in the matter
sector (see \cite{9} for a review). From the complete set of
one-loop divergences, given explicitly in Ref.~
\cite{3}, there are no problems in writing down all
running coupling constants, including $M\neq 0$. To save space we have
listed only those needed in this section.

Working with the massless version of the theory (\ref{2.1}) we use
(\ref{2.2}) first to find the effective potential at one-loop and
second to find  RG improved asymptotic forms of this potential.
We start by writing the effective action of this theory as:
\beq
\Gamma = \left. \Gamma \right|_{\phi =0} + \int d^6x \sqrt{g} \,
V + \cdots, \label{2.4}
\eeq
where the first term is the vacuum energy, the second is the
effective potential. Terms that have not been explicitly included
provide non-constant $\phi$ contributions to $\Gamma$.
The multiplicative renormalizability of the theory guarantees
that the effective action as well as the effective potential
satisfies the RG equations:
\beq
\left( \mu \frac{\partial}{\partial \mu} + \beta_{\lambda_i}
\frac{\partial}{\partial \lambda_i} - \gamma_\phi \phi
\frac{\partial}{\partial \phi} \right) V =0,
\label{2.5}
\eeq
where $\gamma_{\phi} $ is the $\gamma$-function of the scalar field
(computed here from \cite{3}),
\beq
\gamma_\phi={g^2\over 12(4\pi)^3}. \label{2.4+}
\eeq

In order to find the effective potential as an expansion over
curvature invariants, we will write the classical potential as (its form
is clear from (\ref{2.1}))
 \beq
V^{(0)} = \sum_i  V^{(0)}_i, \qquad V^{(0)}_i =
a_i\lambda_iP_i\varphi^{k_i},
\label{2.6}
\eeq
where the $a_i$ are numerical multipliers, $k_i\geq 1$ are integers, and the
$P_i$ are curvature invariants. Applying the method described in
Ref.~\cite{5} (see also \cite{4}), we can solve the RG equations
(\ref{2.5}) for a potential of the form (\ref{2.6}). Restricting
ourselves to one-loop and using the tree level
potential (\ref{2.6}) as boundary condition, we find (we skip
technical details): \bea
V &=& \frac{1}{6} g\phi^3 - \frac{g^3\phi^3}{12(4\pi)^3} \left(
\ln \frac{\phi^2}{\mu^2} - \frac{11}{3} \right)+ h \phi +
\frac{\xi}{2} R\phi^2 - \frac{g^2}{4(4\pi)^3} \left( \xi -
\frac{1}{6} \right) R\phi^2 \left( \ln \frac{\phi^2}{\mu^2} - 3
\right) \nn \\ && + \eta_1 R\phi + \eta_3 R^2 \phi + \eta_4
R_{\mu\nu}^2 \phi +  \eta_5 R_{\mu\nu\alpha\beta}^2 \phi
\label{2.7} \\ && - \frac{g\phi}{(4\pi)^3} \left( \ln
\frac{\phi^2}{\mu^2} - 2 \right) \left[ \frac{1}{4} \left( \xi -
\frac{1}{6} \right)^2 R^2 -\frac{1}{360}R_{\mu\nu}^2
+\frac{1}{360}  R_{\mu\nu\alpha\beta}^2  \right]. \nn
\eea
This is the one-loop effective potential up to
terms quadratic in the curvature. It is clear that this potential
is not bounded from below (this is the
well-known instability of the $\varphi_6^3$-theory). This kind of
potential is useful for studying  six-dimensional cosmology  coupled to  a
$\varphi_6^3$ quantum
field.

Another applications of the RG
equations to the effective potential $V$, is to study the
asymptotics of the effective potential in curved spacetime
\cite{6,4}. The homogeneity condition of $V$
 has the form
\beq
V \left( e^{2t}\phi, e^{d_{\lambda_i}t} \lambda_i, e^{-
2t}g_{\alpha\beta}, e^t\mu \right) = e^{6t} V \left( \phi,
\lambda_i,g_{\alpha\beta},\mu \right) ,
\label{2.8}
\eeq
where $t=$ const. and $d_{\lambda_i}$ is the dimension of
$\lambda_i$. Relation (\ref{2.8}) leads to the following
equations:
\bea
&& \left( \partial_t +\mu\partial_\mu + d_{\lambda_i} \lambda_i
 \partial_{\lambda_i} + 2\phi \partial_\phi -6\right) V \left(
\phi, e^{-2t} g_{\alpha\beta},  \ldots \right) = 0,
\label{2.9} \\
&& \left( \partial_t +\mu\partial_\mu + d_{\lambda_i} \lambda_i
 \partial_{\lambda_i} - 2g_{\alpha\beta} \frac{\delta}{\delta
g_{\alpha\beta}}-6\right) V \left( e^{2t} \phi, g_{\alpha\beta},
\ldots \right) = 0, \label{2.10}
\eea
where the parameters of the potential that are not written explicitly
are not scaled.
Combining Eq.~(\ref{2.5}) with Eqs.~(\ref{2.9}) and (\ref{2.10}),
we obtain the following:
\bea
&& \left[ \partial_t - \left( \beta_{\lambda_i} - d_{\lambda_i}
\lambda_i \right)  \partial_{\lambda_i} + \left( \gamma_\phi + 2
\right) \phi \partial_\phi -6\right] V \left( \phi,
e^{-2t} g_{\alpha\beta},  \ldots \right) = 0, \label{2.11} \\
&& \hspace{-1cm} \left\{ \partial_t + ( 1 + \gamma_\phi /2)^{-1} \left[ -
\left( \beta_{\lambda_i} - d_{\lambda_i} \lambda_i \right)
\partial_{\lambda_i} - 2g_{\alpha\beta} \frac{\delta}{\delta
g_{\alpha\beta}}-6\right] \right\} V \left( e^{2t} \phi,
g_{\alpha\beta}, \ldots \right) = 0. \label{2.12}
\eea
The RG equations (\ref{2.11}) and (\ref{2.12})  describe the
asymptotics of the effective potential. In particular, when
$g_{\alpha\beta} \rightarrow e^{-2t} g_{\alpha\beta}$, $R^2
\rightarrow e^{4t} R^2$, Eq.~(\ref{2.11}) gives the
asymptotic behavior of the effective potential in a strong gravitational
field. Similarly, Eq.~(\ref{2.12}) gives the behavior of $V$ in the
case of a strong scalar field. Solving Eq.~(\ref{2.11}) we get (see
also \cite{6,4}):
\bea
&& V \left( \phi, e^{-2t}  g_{\alpha\beta}, \lambda_i \right) =
e^{6t}  V \left( \phi (t),  g_{\alpha\beta}, \lambda_i (t) \right),
\nn \\
&& \dot{\lambda}_i (t) = \beta_{\lambda_i} (t) - d_{\lambda_i}
\lambda_i (t), \qquad \lambda_i (0) = \lambda_i, \label{2.13} \\
&& \dot{\phi} (t) = - \left[ 2 + \gamma_\phi (t) \right] \phi (t),
\qquad \phi (0) = \phi. \nn
\eea
Selecting the leading coupling constants from
(\ref{2.2}) and using (\ref{2.4+})
we obtain
\beq
 V \left( \phi, e^{-2t}  g_{\alpha\beta}, \lambda_i \right) \sim
e^{6t} \phi (t) \left[ \eta_3(t) R^2 + \eta_4(t) R_{\mu\nu}^2 +
\eta_5(t) R_{\mu\nu\alpha\beta}^2 \right], \label{2.15}
\eeq
where
\beq
\phi (t) = \phi e^{-2t}B^{-1/18} (t). \label{2.14}
\eeq
Thus, the asymptotics of the effective potential in a strong
gravitational field are defined by the non-minimal interaction of
the scalar with the quadratic curvature invariants. Such
approximations can be useful in studying quantum effects in the
early universe (e.g. in the Kaluza-Klein framework).

In a similar way, we can solve Eq.~(\ref{2.12}), with the result
\beq
 V \left( e^{2t} \phi, g_{\alpha\beta}, \lambda_i \right) =
\exp \left[ 6\int_0^t dt'\, A(t') \right] \,
V \left( \phi ,  g_{\alpha\beta} (t), \wt{\lambda}_i (t)
\right), \label{2.16}
\eeq
where
\bea
&& A(t) = \left[ 1 + \frac{ \gamma_\phi (t)}{2} \right]^{-1},
\qquad \phi (t) = \phi, \nn \\
&& \dot{g}_{\alpha\beta} (t,x) = 2A(t) g_{\alpha\beta} (t,x),
\qquad g_{\alpha\beta} (0,x) = g_{\alpha\beta} (x),  \nn \\
&& \dot{\wt{\lambda}}_i (t) = A(t) \left[ \beta_{\wt{\lambda}_i}
(t) - d_{\wt{\lambda}_i} \wt{\lambda}_i (t) \right].
 \label{2.17}
\eea
As we see, contrary to what happens with Eqs.~(\ref{2.13}) for the effective
couplings, the multiplier
$A(t)$ appears on the
rhs of Eqs.~(\ref{2.17}). Using arguments similar to the ones given in Ref.~
\cite{7} (where the procedure to study the asymptotics of the
effective potential in flat spacetime was developed), one can show
that the presence of $A(t)$ does not influence the asymptotics of
the effective couplings. Again, due to the fact that the theory is
asymptotically free, it is natural to expect  that the asymptotic
behavior of the effective potential is given by the lowest order of
perturbation theory, with the parameters replaced by the
corresponding effective couplings.

Now, since $\phi (t)=\phi$, and the effective curvature is always
small, $R(t) \sim e^{-2t}$ (see (\ref{2.17})), we get
\beq
 V \left( e^{2t} \phi, g_{\alpha\beta}, \lambda_i \right) =
\frac{1}{6} e^{6t} g(t) \phi^3.
 \label{2.18}
\eeq
The asymptotic value of the effective potential, in the limit of strong
scalar curvature,  is not bounded from below. This result can be useful for the study of
six-dimensional quantum cosmology near the initial singularity.
We conclude this discussion of the application of RG equations to the effective potential for the curved spacetime
$\phi_6^3$-theory and go on to an application the RG equations  to the vacuum energy.
\bs


\bce
{\large \bf III. \ The vacuum energy in the $\phi_6^3$-theory on a
Kaluza-Klein spacetime.}
\ece

Starting from the works \cite{12} and \cite{8}, the vacuum
energy of matter and gravitational fields on spherically
compactified internal spaces was calculated and the process of quantum
spontaneous compactification was studied. For a review and a list of references
of papers on
related question concerning Kaluza-Klein theories, see \cite{10,9}.
In particular, in \cite{13}-\cite{17} and \cite{11} vacuum
energies were evaluated for scalar fields etc. (including gravity) defined on
even-dimensional compactified spaces. In most of these studies
only the divergent parts (in dim-reg) of the vacuum energies were evaluated.

Our goal here is to obtain the RG improved one-loop vacuum energies corresponding to
the theory (\ref{2.1}) on two Kaluza-Klein backgrounds, namely
$R^4\times S^1\times S^1$ and $R^4\times S^2$, and to investigate
the process of spontaneous compactification.
\ms

\ni {\bf (a) $R^4\times S^1\times S^1$ space.}

At the one-loop level, the vacuum energy
is given by
\beq
\Gamma^{(1)} = \frac{1}{2} \tr \ln \left( - \Box + M^2 \right).
 \label{2.19}
\eeq
The calculation can be done with the help of zeta function
regularization (for an introduction, see \cite{18}).
The spectrum has the form
\beq
\lambda = k_4^2 + \left( \frac{2\pi n_1}{L_1} \right)^2 +
 \left( \frac{2\pi n_2}{L_2} \right)^2 + X,
 \label{2.20}
\eeq
with $X=M^2$ here, and the corresponding  `Euclideanized' zeta function is
\bea
\zeta_E (s) &=& \sum_\lambda \lambda^{-s} = \frac{1}{\Gamma (s)}
\sum_\lambda \int_0^\infty dt \, t^{s-1} e^{-\lambda t} \nn \\
&=&  \frac{1}{\Gamma (s)} \sum_{n_1n_2} \int \frac{d^4k}{(2\pi)^4}
\int_0^\infty dt \, t^{s-1} \exp \left\{-
\left[ k^2 + \left( \frac{2\pi n_1}{L_1} \right)^2 +
 \left( \frac{2\pi n_2}{L_2} \right)^2 + X\right] t\right\} \nn \\
&=& \frac{1}{4(2\pi)^{2(s+1)} (s-1)(s-2)} \left[
\left( \frac{X}{4\pi^2} \right)^{2-s} + E \left( s-2;L_1^{-2}, 0,
L_2^{-2}; \frac{X}{4\pi^2} \right) \right].
 \label{2.21}
\eea
$E(s;a,b,c;q)$ is the zeta function introduced and studied in
\cite{Eli1},
\beq
E(s;a,b,c;q) \equiv
{\sum_{m,n \in \mbox{\bf Z}}}' (am^2+bmn+cn^2+q)^{-s},
\quad \mbox{Re} \, (s) >1.
\label{2.22}
\eeq
In the general theory \cite{Eli1}, one requires that $a,c >0$,  that
the discriminant
\beq
\Delta =4ac-b^2 = \left( \frac{2}{L_1 L_2} \right)^2 >0,
\label{2.23}
\eeq
and that $am^2+bmn+cn^2+q \neq 0$, for all $ m,n \in
$ {\bf Z}.
These
conditions are all satisfied in this case.
The analytic continuation \cite{Eli2} of this zeta function is:
\bea
&& \hspace{-5mm} E(s;a,b,c;q) = -q^{-s}
+\frac{2\pi q^{1-s}}{(s-1) \sqrt{\Delta}}
 + \frac{4}{\Gamma (s)} \left[
\left( \frac{q}{a} \right)^{1/4}
 \left( \frac{\pi}{\sqrt{qa}} \right)^s
\sum_{k=1}^\infty
k^{s-1/2} K_{s-1/2} \left( 2\pi k \sqrt{\frac{q}{a}} \right) \right.
\nn \\ &&  \hspace{6cm} + \sqrt{\frac{q}{a}} \left(2\pi
\sqrt{\frac{a}{q\Delta}} \right)^s
 \sum_{k=1}^\infty k^{s-1} K_{s-1} \left( 4\pi k
\sqrt{\frac{a q}{\Delta}}\right)  \label{cse1}
 \\ && +\left. \sqrt{\frac{2}{a}} (2\pi)^s
 \sum_{k=1}^\infty k^{s-1/2} \cos (\pi k b/a) \sum_{d|k} d^{1-2s} \,
 \left( \Delta + \frac{4aq}{d^2} \right)^{1/4-s/2}
K_{s-1/2}  \left( \frac{\pi k}{a} \sqrt{ \Delta + \frac{4aq}{d^2}}
\right) \right].\nn
\eea
This explicit form
(\ref{cse1})
 and its
derivative (given below)  appeared for the first time in  \cite{Eli2}.
It is remarkable that the only simple pole ($s=1$) is so explicit in (\ref{cse1}).
This expression also has excellent convergence
properties, in fact, for
large $q$ the convergence behavior of the series of Bessel functions
is at least exponential.
Particular values for
 $s=-n$, $n=0,1,2,3,\ldots$ are :
 \beq
 E(-n;a,b,c;q) = -q^n - \frac{2\pi}{n+1} \,
\frac{q^{n+1}}{\sqrt{\Delta}},
\eeq
and
 \beq
 E(-n;a,b,c;0) =0.
\eeq
For the corresponding derivative at zero we have
\bea
E'(0;a,b,c;q) &=& -\frac{2\pi q}{\sqrt{\Delta}}+ \left( 1 +
\frac{2\pi q}{\sqrt{\Delta}} \right) \ln q \nn \\
&&-2 \ln \left( 1 - e^{-2\pi \sqrt{q/a}} \right)
+ 4 \sqrt{\frac{q}{a}} \sum_{n=1}^\infty n^{-1}  K_1 \left( 4n\pi
\sqrt{\frac{aq}{\Delta}} \right)
\label{eh22n} \\
&& + 4\sum_{n=1}^\infty
n^{-1} \cos (n\pi b/a) \sum_{d|n} d \,
\exp \left[
-\frac{\pi n}{a} \left( \Delta + \frac{4aq}{d^2}
\right)^{1/2} \right],    \nn
\eea
and, in general, for  $s=-n$, $n=0,1,2,3,\ldots$,
\bea
&&\hspace{-5mm} E'(-n;a,b,c;q) =
-\frac{2\pi q^{n+1}}{(n+1)^2 \sqrt{\Delta}}+
q^n \left( 1 + \frac{2\pi q}{(n+1)\sqrt{\Delta}} \right) \ln q \nn \\
&&  \hspace{1cm}
+ 4 \frac{(-1)^n}{n!\, \pi^n} \left[ q^{n/2+1/4}a^{n/2-1/4}
\sum_{k=1}^\infty
k^{-n-1/2} K_{n+1/2} \left( 2\pi k \sqrt{\frac{q}{a}} \right) \right.
\nn \\ &&  \hspace{3cm} + 2^{-n} \left(\frac{q}{a} \right)^{(n+1)/2}
\Delta^{n/2} \sum_{k=1}^\infty k^{-n-1} K_{n+1} \left( 4\pi k
\sqrt{\frac{a q}{\Delta}} \right)
 \\ && + \left. \frac{2^{-n+1/2}}{\sqrt{a}}
 \sum_{k=1}^\infty k^{-n-1/2} \cos (\pi k b/a) \sum_{d|k} d^{2n+1} \,
 \left( \Delta + \frac{4aq}{d^2} \right)^{k/2+1/4}
K_{n+1/2}
 \left( \frac{\pi k}{a} \sqrt{ \Delta + \frac{4aq}{d^2}}
\right) \right]. \nn
\eea
These are the only expressions needed for what follows.
We want to evaluate the effective action
$\Gamma^{(1)} /V_4$, where
\beq
\Gamma^{(1)} = \frac{1}{2} \left[ {\zeta'}_E (0) + \zeta_E (0) \ln
\mu^2 \right],
\label{2.24}
\eeq
and $V_4$ is the four-volume, $V_4 \equiv \int d^4x$.
The result is immediate from the expressions above:
\bea
\frac{\Gamma^{(1)}}{V_4} &=& \frac{M^6L_1L_2}{128 \pi^3} \left(
-\frac{11}{36} + \frac{1}{6} \ln \frac{M^2}{\mu^2} \right) +
2 \sqrt{\frac{2}{\pi}} \, \frac{M^{5/2}}{L_1^{3/2}} \sum_{n=1}^\infty
n^{-5/2} K_{5/2} \left( nML_1\right) \nn \\
&& + \frac{M^3L_1}{4\pi  L_2^2} \sum_{n=1}^\infty n^{-3} K_3
 \left( nML_1 \right) \label{2.25} \\
&& +
\frac{2\sqrt{2}\, \pi^2}{L_1^{3/2}} \sum_{n=1}^\infty n^{-5/2}
\sum_{d|n} d^5 \left(\frac{4}{L_2^2} + \frac{M^2}{\pi^2d^2}
\right)^{5/4} K_{5/2}  \left( \pi nL_1 \sqrt{
\frac{4}{L_2^2} + \frac{M^2}{\pi^2d^2} }\right).\nn
\eea
Notice that the result is given in terms of a rapidly convergent
series, very well suited for numerical computation. In the massless case
($M^2 =0$), we are left with the last term
\beq
\left. \frac{\Gamma^{(1)}}{V_4} \right|_{M^2=0}  =
\frac{16\, \pi^2}{L_1^{3/2}L_2^{5/2}} \sum_{n=1}^\infty n^{-5/2}
\sigma_5 (n) \, K_{5/2}  \left( 2\pi n \frac{L_1}{L_2} \right).
\eeq

\ms

\ni {\bf (b) $R^4\times S^2$ space.}

In this case, for simplicity,
the vacuum energy will be calculated for the massless theory only
\beq
\Gamma^{(1)} = \frac{1}{2} \tr \ln \left( - \Box + \xi R \right).
 \label{3b1}
\eeq
The spectrum is now
\beq
\lambda = k_4^2 - \Lambda_l^2 + X,
 \label{3b2}
\eeq
where $X=\xi R$. For the 2-sphere $ R=2/r^2$ when written in terms of the
sphere's radius $r$. For scalar fields,
\beq
\Lambda_l^2 = - \frac{l(l+1)}{r^2}, \qquad l = 0,1,2,\ldots
 \label{3b3}
\eeq
with associated multiplicities
\beq
D_l =  2l+1.
 \label{3b4}
\eeq
The corresponding zeta function is
\bea
\zeta_E (s) &=& \frac{\Gamma (s-2)}{16\pi^2 \Gamma (s)}
\sum_l D_l (\Lambda_l^2 +X)^{2-s} \nn \\
&=&  \frac{r^{2(s-2)}}{16\pi^2(s-1)(s-2)} \sum_{l=0}^\infty
(2l+1) \left[ ( l + 1/2)^2 + (Xr^2 -1/4) \right]^{2-s} \label{3b5}
\\ &=& -\frac{r^{2(s-2)}}{16\pi^2(s-1)(s-2)(s-3)}
\left. \frac{\partial}{\partial c} F(s-3;c;Xr^2-1/4) \right|_{c=1/2},
\nn \eea
where $F(s;c;q)$ is another typical zeta function  studied in full detail in
\cite{Eli1},
\beq
F(s;c;q) \equiv \sum_{n=0}^\infty \left[ (n+c)^2 +q\right]^{-s} \equiv
G(s;1,c;q).
\label{3b6}
\eeq
 From the general asymptotic expansion of
 $G(s;a,c;q)$ in powers of $q^{-1}$ (see \cite{Eli1}),
\bea
&& \hspace{-8mm} G(s;a,c;q) \equiv \sum_{n=0}^{\infty} \left[ a(n+c)^2+q
\right]^{-s} \sim \frac{q^{-s}}{\Gamma (s)} \sum_{m=0}^{\infty}
\frac{(-1)^m \Gamma (m+s)}{m!} \left( \frac{q}{a} \right)^{-m}
\zeta_H (-2m, c) \ \ \ \ \label{if11} \\ &&   +
\sqrt{\frac{\pi}{a}} \, \frac{\Gamma (s-1/2)}{2\Gamma (s)} q^{1/2 -s}
+\frac{2\pi^s}{\Gamma (s)} a^{-1/4-s/2} q^{1/4-s/2}
 \sum_{n=1}^\infty
n^{s-1/2} \cos (2\pi nc) K_{s-1/2} (2\pi n\sqrt{q/a}), \nn
\eea
we easily obtain the asymptotic expansion:
\beq
\zeta_E (s) \sim -\frac{r^{2(s-2)}}{16\pi^2} \sum_{n=0}^\infty
\frac{(-1)^n \left(1-2^{1-2n} \right) B_{2n} \Gamma (s+n-3)}{n! \,
\Gamma (s)} \, (Xr^2-1/4)^{3-s-n},
\label{3b7}
\eeq
where the $B_{2n}$ are Bernoulli numbers. This yields immediately
\bea
\zeta_E (0)&=& \frac{1}{16\pi^2r^4} (Xr^2-1/4)^3
\left[ - \frac{1}{6} +
\frac{1}{24}  (Xr^2-1/4)^{-1}
-\frac{7}{480}  (Xr^2-1/4)^{-2} \right. \nn \\ && \hspace{5cm} \left.
+\frac{31}{8064}  (Xr^2-1/4)^{-3} \right],
\label{3b8}
\eea
and
\bea
{\zeta'}_E (0)&=& -\zeta_E(0) \ln \left( X -\frac{1}{4r^2} \right)
+\frac{1}{16\pi^2r^4} (Xr^2-1/4)^3 \left[ - \frac{11}{36}
+ \frac{1}{16}  (Xr^2-1/4)^{-1} \right. \nn \\ && \left.
-\frac{7}{480}  (Xr^2-1/4)^{-2}
+\sum_{n=4}^\infty
\frac{(-1)^{n+1} \left(1-2^{1-2n} \right) B_{2n}}{n(n-1)(n-2)(n-3)
} \, (Xr^2-1/4)^{-n} \right].
\label{3b9}
\eea
Finally,
\bea
\frac{\Gamma^{(1)}}{V_4} &=& \frac{1}{32 \pi^2 r^4} \left(
 2\xi -\frac{1}{4} \right)^3 \left\{ \left[ -\frac{11}{36}
+ \frac{1}{16}  (2\xi-1/4)^{-1} \right. \right. \nn \\ && \left.
-\frac{7}{480}  (2\xi-1/4)^{-2}
+\sum_{n=4}^\infty
\frac{(-1)^{n+1} \left(1-2^{1-2n} \right) B_{2n}}{n(n-1)(n-2)(n-3)
} \, (2\xi-1/4)^{-n} \right] \label{3b10} \\ &&  \hspace{-2cm} + \left.
\ln \left( \frac{\mu^2r^2}{2\xi - 1/4} \right)
\left[ - \frac{1}{6} +
\frac{1}{24}  (2\xi-1/4)^{-1}
-\frac{7}{480}  (2\xi-1/4)^{-2}
+\frac{31}{8064}  (2\xi-1/4)^{-3} \right] \right\}, \nn
\eea
which is an asymptotic expansion for large $\xi$, valid for $\xi > 1/8$.
The optimal truncation of this asymptotic expansion is obtained
after the $n=4$ term.
The point $\xi = 1/8$ has nothing to do with the conformal
coupling value but instead  depends on how the expansion of $\zeta_E (s)$
was done. This result ceases to be valid when $\xi \leq 1/8$.
For the particular value  $\xi =
1/8$, we can go back to the definition of the zeta function and show that $F$
reduces to an ordinary Riemann zeta
function: $F(s;1/2;0) = \zeta_H (2s;1/2) = (2^{2s}-1) \zeta (2s)$, and
$ \frac{\partial}{\partial c} \zeta_H (s;c) =-s\zeta_H (s+1;c)$,
resulting in
\beq
\zeta_E (s) =
\frac{r^{2(s-2)}(2^{2s-5}-1)}{8\pi^2(s-1)(s-2)} \zeta (2s-5), \qquad \xi
= 1/8.
 \eeq
Then
\beq
\frac{\Gamma^{(1)}}{V_4} = \frac{1}{2^{11} \pi^3 r^4} \left\{
\frac{31}{504} \left[ \frac{3}{2} + \ln \left(\mu^2r^2\right) \right]
- \frac{\ln 2}{252} - 31 \zeta'(-5) \right\}, \qquad  \xi =1/8.
\eeq
For $\xi < 1/8$ the expansion (\ref{3b10}) is
replaced by a convergent series stemming
from the binomial expansion  of $F(s;,c;q)$ in powers of $q$, $0 \leq
q <1$  (see Eq.~(\ref{3b6})). This can be easily done, as
described in detail in \cite{Eli1,Eli2}. One would then have
expressions which cover the whole range of values of $\xi$.
However, in order to limit discussion we
will  illustrate the physical argument with the help of the $\xi >1/8$
case, i.e., Eq.~(\ref{3b10}). Similar considerations would also apply
 to the $\xi \leq 1/8$ cases.
\ms

\ni {\bf (c) Spontaneous compactification.}

Using the vacuum energies that we have calculated above, we can now study
the process of quantum spontaneous compactification on  Kaluza-Klein
backgrounds (see, e.g. \cite{10}).
In particular, we want to investigate
consequences of using RG improvements to these energies on spontaneous
compactification.

Turning back to our first example, we will now consider a $\phi_6^3$
theory on a $R^4\times S^1\times S^1$ background where, for simplicity,
we set $L_1=L_2= L$. The effective action which takes into account the
one-loop corrections (\ref{2.25}) is given by
\beq
\frac{\Gamma}{V_4} = - \Lambda L^2 + \frac{\Gamma^{(1)}}{V_4}.
\label{2.41a}
\eeq
The conditions of spontaneous compactification are:
\beq
\Gamma =0, \qquad \frac{\partial}{\partial L}
\left( \frac{\Gamma}{V_4} \right) = 0.
\label{2.42}
\eeq
In the case under discussion, we have  $\Gamma = \Gamma (\Lambda,
M^2,\mu, L)$. Having two conditions and four parameters, the
expectations of finding some solution of Eq. (\ref{2.42}) are great.
We will fix $M^2$ and $\mu$ and consider $\Gamma$ as a function of the
compactification length, $L$, only. In Fig. 1 we  call this
effective action (i.e., $\Gamma$ divided by the four-volume) simply
$V(L)$, and show its form
explicitly for some specified values of $M^2, \mu$ and $\Lambda$ that
correspond to one of those situations in which spontaneous
compactification takes place for a definite value of the
compactification length $L$. Note that
in this case we are beyond the range of validity of our approximation
(which is analogous to that of the Coleman-Weinberg potential
\cite{22}), because of the large logarithmic contribution. For
reasonably
small values of the log (where the one-loop result can be trusted), that
is up to  $|\ln (M^2 /\mu^2 )| \simeq 1$, there is no minimum.

Let us now see how this picture changes, in general, when we take into
account RG effects, e.g., when we enlarge the parameter space.
 As the theory is multiplicatively renormalizable, the
effective action satisfies the RG equation (\ref{2.5}). This equation
can be solved using the method of the characteristics, yielding the
so-called RG improved effective action (or Wilsonian effective action
\cite{21}). The corresponding RG improved effective potential \cite{22}
has been widely discussed in renormalizable theories with a Higgs
sector, both in flat  \cite{23} and in curved spacetime \cite{24}.

The solution of the RG equation (\ref{2.5}) (at
$\phi =0$) for the effective action gives
\beq
\Gamma (\lambda_i, g_{\alpha\beta}, \mu) = \Gamma (\lambda_i (t),
 g_{\alpha\beta}, \mu e^t ),
\label{2.43}
\eeq
where the effective couplings are given in (\ref{2.2}). As boundary
condition for (\ref{2.43}) it is convenient to use the one-loop
effective action (\ref{2.41a}). Then, the RG improved effective action
is
given by the same expression (\ref{2.41a}) but with the following
changes of variables:
\bea
&& M^2  \longrightarrow  M^2(t) = M^2 \left( 1 + \frac{3g^2t}{2(4\pi)^3}
\right)^{-5/9}, \nn \\
&& \mu^2  \longrightarrow  \mu^2(t) = \mu^2 e^{2t}, \qquad
\Lambda  \longrightarrow  \Lambda (t) = \Lambda -\frac{ M^2}{6g^2}
\left[ \left( 1 + \frac{3g^2t}{2(4\pi)^3}
\right)^{-2/3}- 1 \right]. \label{2.44}
\eea
In order to define $t$ we may choose the standard and most natural
condition of dropping out the logarithmic term (for more details and
different ways of defining $t$, see the last reference in \cite{23})
\beq M^2(t)= \mu^2 e^{2t}. \label{2.45}
\eeq
The solution of (\ref{2.45}) determines the value of $t$ as a
function of $g$, $M^2$ and $\mu$.
Fixing
$M^2$ and $\mu$, as before, we obtain now the corresponding picture for
the RG improved effective potential $\Gamma$ as a function of $L$. This
is depicted in Fig. 2 for specified values of $M,\mu,\Lambda$ and
$g$ that correspond to those of Fig. 1. The value of $t$ which is a
solution of Eq. (\ref{2.45})
is here $t=3.2182$. Differences in the effective potential $\Gamma$ as a
function of $L$ caused by the RG improvement can be seen by
comparing Figs. 1 and 2. A virtue of the RG improvement is that now it
has physical sense to let the quocient $M^2/\mu^2$  take such big
values as in Fig. 1: the range of validity of the approximation is
greatly enlarged. However, as is clearly
observed in these figures, the RG improvement can modify the spontaneous
compactification pattern dramatically. It actually happens in this
case: the minimum disappears completely, at best it can turn into
an inflection point (of the type of Fig. 3, although here it
corresponds to the non-improved case). We see also that the
renormalization group acting over an inflection point makes it evolve
into the non-compactifying
case (cf. Figs. 3 and 4).
 And this is all what happens in the model considered:
the effect of the RG is destructive, concerning spontaneous
compactification (it need not be so in other situations).

The same calculation can be repeated for the case of the RG improved
effective action corresponding to the theory defined on the space
$R^4\times S^2$. Using the same principle as above, we can write
\beq
\frac{\Gamma}{V_4} = \frac{1}{V_4} \int d^6 x \, \sqrt{g} \, L_{ext} +
\frac{\Gamma^{(1)}}{V_4},
\label{2.48}
\eeq
where $L_{ext} = L_{ext} (\Lambda, k, \alpha_1, \ldots, \alpha_{10})$
(see Eq. (\ref{2.1})), and $\Gamma^{(1)}$ is given by (\ref{3b10}).
Expression (\ref{2.48}) yields the one-loop effective action.
The corresponding RG improved effective action is given by the same
formula (\ref{2.48}) but with the substitutions:
\beq
\xi \longrightarrow  \xi (t) = \frac{1}{5} + \left( 1- \frac{1}{5}
\right) B^{-5/9} (t), \qquad
\alpha_i \longrightarrow  \alpha_i (t),
\eeq
where $i=4,5, \ldots, 10$. The explicit form of $\alpha_i (t)$ can be
easily obtained from Refs. \cite{3,4}. One can establish the same
comparison as before between the results of the
spontaneous compactification process corresponding to the one-loop and
to the RG improved effective actions on $R^4\times S^2$. In
 this case
there are a total of 14 parameters to satisfy just two conditions.
Setting aside exceptional situations, an enormous
variety of possibilities occur. For this case the
difference in the process of compactification introduced by the RG
improvement of the effective action is found again.
We omit  plots similar to ones corresponding
to the previous example.
\bs

\bce
{\large \bf IV. \ Conclusions.}
\ece

In this work we have investigated renormalization group effects in the $\phi_6^3$
curved
spacetime theory. Using this case as an example, the
usefulness of the RG improvement in higher-dimensional theories has been
demonstrated by calculating the one-loop effective potential and
its asymptotics in  strong fields. Additionally using the one-loop
vacuum energy and the RG improved vacuum energy
on Kaluza-Klein backgrounds $R^4\times S^1 \times S^1$ and
$R^4\times S^2$, we have shown that the RG improvement can lead to
significant qualitative differences in the spontaneous compactification of these
theories.
Such results clearly show that, in general, one-loop predictions in
Kaluza-Klein theories should not always be trusted.

There are not many theories which are renormalizable in the standard
way in higher dimensions. But the number increases by
the introduction of higher-derivative kinetic terms (probably paying the price
of spoiling unitary). For example, in
$D=6$ one can consider a gauge theory with $F_{\mu\nu}^3$-terms as
kinetic terms plus any other term not prohibited by dimensional
arguments (they can have equal or lower dimensionality) or gauge
invariance. Such a theory will be renormalizable in the same sense as $R^2$-gravity
in $D=4$  is (see \cite{9} for a review).

On the other hand, one can consider Kaluza-Klein theories to be
renormalizable in the sense of the inclusion of an infinite number of
additional terms and their corresponding counterterms (for a recent
discussion, see \cite{25}). Under these circumstances,
 the RG analysis can
be applied again, but, of course, the RG equations will be
infinite in number. Nevertheless, there are ways of
truncating them in a systematic and consistent way,
 keeping just a finite number of terms. This can be done by considering,
say just one-loop effects (in the even-dimensional case), or terms up
to some particular order in the derivatives.
As we have shown here, if the goal is studying spontaneous
quantum compactification, the RG improved vacuum
energy should  be used.


\vspace{5mm}

\noindent{\bf Acknowledgments.}
SDO is grateful to
the members of the Department ECM, Barcelona University, for warm
hospitality.
This work has been supported by DGICYT (Spain), project
PB93-0035 and grant SAB93-0024, by CIRIT (Generalitat de
Catalunya), grant GR94-8001, and by RFFR, project 94-020324. RK is
supported by the US Department of Energy.

\newpage

\newpage

\ni{\large \bf Figure captions}
\ms

\ni {\bf Fig. 1:} The effective action $V\equiv \Gamma /V_4$, as a
function
of $L$, for the specified values of $M^2, \mu$ and $\Lambda$ (in units
of $10^4$). They
correspond to a situation in which spontaneous
compactification takes place. At the minimum, the value of the
compactification length $L$ is selected.
 \ms

\ni {\bf Fig. 2:} Plot of the renormalization group improved effective
action for values of $M,\mu,\Lambda$ and
$g$ that correspond to those of Fig. 1.
When comparing with Fig. 1, it is clearly observed that
the RG improvement can modify the spontaneous
compactification pattern.
 \ms

\ni {\bf Fig. 3:} Plot of the effective action
 for values of $M,\mu,\Lambda$ and
$g$ that yield an inflection point, a situation that stays in the verge
of spontaneous compactification.

 \ms

\ni {\bf Fig. 4:} Plot of the RG improved effective
action for values of $M,\mu,\Lambda$ and
$g$ corresponding to Fig. 3. The inflection point has disappeared
and we are driven far away from spontaneous compactification.

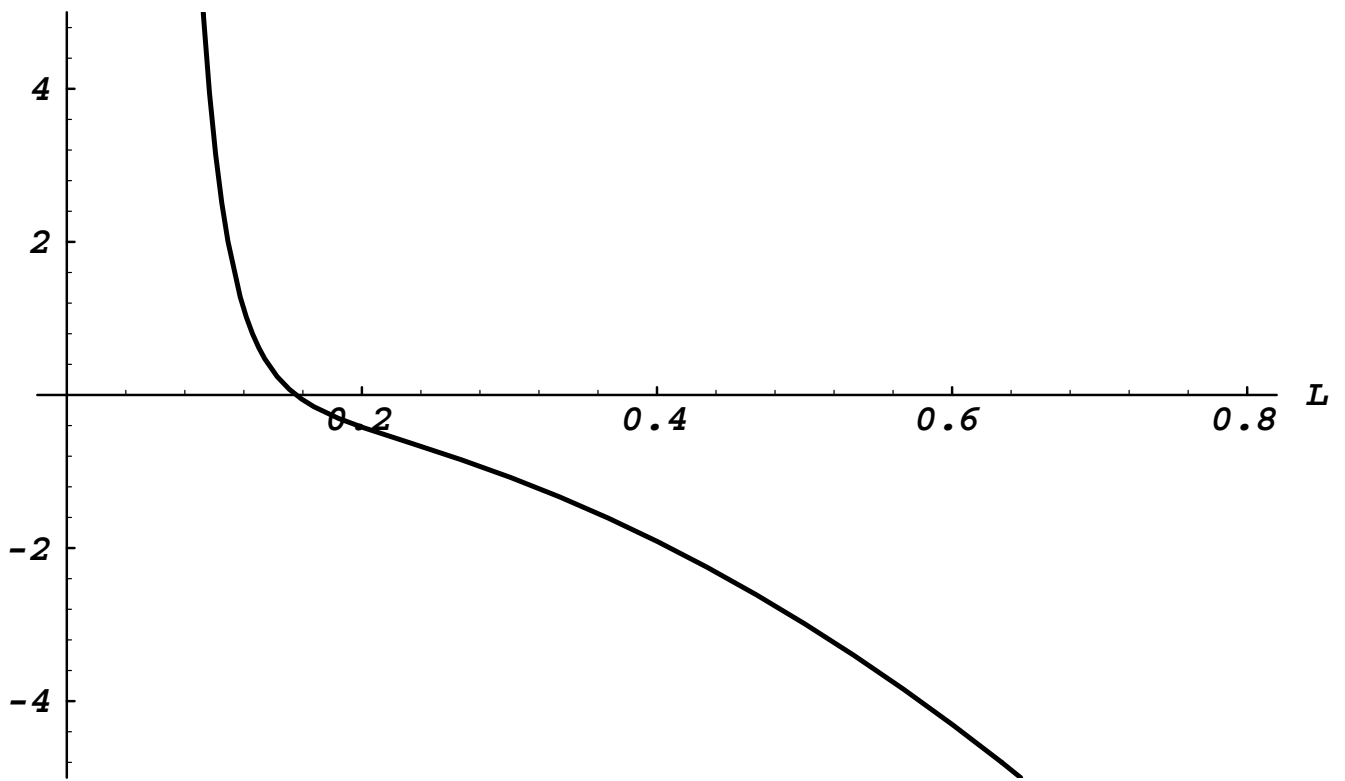
\end{document}